\documentstyle[preprint,aps,epsf]{revtex}

\begin{document}

\title{Spin polarized neutron matter and magnetic susceptibility within
the Brueckner--Hartree--Fock approximation}

\author{I.\ Vida\~na$^{1,2}$, A.\ Polls$^{2}$ and A.\ Ramos$^{2}$}

\address{$^{1}$Dipartimento di Fisica, Universit\`a di Pisa and INFN Sezione di Pisa, 
Via Buonarroti 2, 56127 Pisa, Italy}
\address{$^{2}$Departament d'Estructura i Constituents de la Mat\`eria,
Universitat de Barcelona, E-08028 Barcelona, Spain}

\maketitle

\begin{abstract}

The  Brueckner--Hartree--Fock formalism is applied to study spin
polarized
neutron matter properties. Results of the total energy per particle as a function of the spin
polarization and density are presented for two modern realistic nucleon-nucleon
interactions, Nijmegen II and Reid93. We find that the dependence of the energy on
the spin polarization is practically parabolic in the full range of polarizations.
The magnetic susceptibility of the
system is  computed. Our results show no indication of a ferromagnetic transition
which becomes even more difficult as the density increases.  


\end{abstract}

\section{Introduction}
\label{sec:sec1}

The study of the magnetic properties of dense matter is of considerable interest in connection 
with the physics of pulsars. These objects, since the suggestion of Gold \cite{go68}, are
generally believed to be rapidly rotating neutron stars with strong surface magnetic fields of
the order of $10^{12}$ Gauss. Several authors have studied the possible existence of a phase
transition to a ferromagnetic state on pure neutron systems at densities corresponding to the 
theoretically stable neutron stars. Brownell and Callaway \cite{br69}, and Rice \cite{ri69}
considered a hard sphere gas model and showed that the ground state of the neutron gas becomes
ferromagnetic at $k_F \approx 2.3$ fm$^{-1}$. Silverstein \cite{si63} and \O stgaard \cite{os70} 
found that the inclusion of long range attraction significantly increased the ferromagnetic
transition density ({\it e.g.} \O stgaard predicted the transition to occur at 
$k_F \approx 4.1$ fm$^{-1}$ using a simple central potential with hard core only for singlet
spin states). Clark \cite{cl69} and Pearson and Saunier \cite{pe70} calculated 
 the magnetic susceptibility for low densities
 ($k_F \leq 2$ fm$^{-1}$)
 using more realistic interactions. 
Pandharipande {\it et al.} \cite{pa72}, using the Reid soft-core potential, performed a
variational calculation arriving to the conclusion that such a transition was not to be 
expected for $k_F \leq 5$ fm$^{-1}$. Early calculations of the magnetic susceptibility
within the Brueckner theory were performed by B\"{a}ckmann and K\"{a}llman \cite{ba73}
employing the Reid soft--core potential, and results from a Correlated Basis Function
calculation were obtained by Jackson {\it et al.} \cite{ja82} with the Reid $v_6$
interaction. A different point of view was followed by Vidaurre {\it et al.} \cite{vi84}, 
who employed neutron-neutron effective interactions of Skyrme type, finding the
ferromagnetic transition at $k_F \approx 1.73$--$1.97$ fm$^{-1}$. 

In connection with the problem of the  neutrino diffusion in dense matter, 
Fantoni {\it et al.} \cite{fa01} have recently employed a new quantum simulation technique (the
so-called Auxiliary Field Diffusion Monte Carlo method (AFDMC)) using realistic interactions
(based upon the Argonne $v_{18}$ two--body potential \cite{sm97} plus Urbana IX three--body
potential \cite{pi98}), and have
found that the magnetic susceptibility of neutron matter shows a strong reduction of about 
a factor $3$ with respect its Fermi gas value. They pointed out that such a reduction may have
strong effects on the mean free path of a neutrino in dense matter and, therefore, it should be 
taken into account in the studies of supernovae and proto-neutron stars. 

In this work we employ the Brueckner--Hartree--Fock (BHF) approximation,
using the realistic 
Nijmegen II and Reid93 \cite{st94} nucleon-nucleon interacions, to study spin polarized 
neutron matter properties such as the total energy per particle and the magnetic
susceptibility. We employ the so-called continuous prescription when solving the 
Bethe--Goldstone equation. As shown by Song {\it  et al.}
\cite{so98}, the effects from three-body clusters are dimished in  this prescription. We also 
explore in this work the dependence of the total energy per particle on the spin
polarization, finding that up to a very good  approximation
this dependence is parabolic.

The paper is organized in the following way. In Section \ref{sec:sec2} the theoretical 
background of  our calculation  is briefly reviewed. The construction
of the neutron-neutron $G$-matrices and the calculation of the total
energy per particle are shown in Section \ref{sec:sec2.1}, whereas the magnetic
susceptibility is determined in Section \ref{sec:sec2.2}. Our results are presented in
Section \ref{sec:sec3}. Finally, a short summary and the main conclusions are given in
Section \ref{sec:sec4}.

%%%%%%%%%%%%%%%%%%%%%%%%%%%%%%%%%%%%%%%%%%%%%%%%%%%%%%%%%%%%%%%%%%%%%%%%%%%%%%%%%%%%%%%%%%%%%%%%%%%%%%%%%%%%%%%%%%%%

\section{Theoretical background}
\label{sec:sec2}

In this section we briefly show how to evaluate, in the BHF approximation, 
the total energy per particle and the magnetic susceptibility of a system of neutrons in which 
we assume that the density of particles with spin up, $\rho^{\uparrow}$, is different from  
that with  spin down, $\rho^{\downarrow}$. 

\subsection{Energy per particle}
\label{sec:sec2.1}

Our calculation of the total energy per particle starts with the construction of 
 $G$-matrices, which describe in an effective way the interaction between
two neutrons, for each one of the spin combinations ($\uparrow\uparrow$,
$\uparrow\downarrow$, 
$\downarrow\uparrow$ or $\downarrow\downarrow$), in the presence of a surrounding medium.
They can be obtained by solving the following integral
Bethe--Goldstone equations

\footnotesize
\begin{eqnarray}
G_{\uparrow\uparrow,\uparrow\uparrow}=
V_{\uparrow\uparrow,\uparrow\uparrow}+V_{\uparrow\uparrow,\uparrow\uparrow}\frac{Q_{\uparrow\uparrow,\uparrow\uparrow}}
{\omega-\epsilon_{\uparrow}-\epsilon_{\uparrow}+i\eta}G_{\uparrow\uparrow,\uparrow\uparrow} \ , \nonumber
\end{eqnarray}

\begin{eqnarray}
G_{\downarrow\downarrow,\downarrow\downarrow}=
V_{\downarrow\downarrow,\downarrow\downarrow}+V_{\downarrow\downarrow,\downarrow\downarrow}
\frac{Q_{\downarrow\downarrow,\downarrow\downarrow}}
{\omega-\epsilon_{\downarrow}-\epsilon_{\downarrow}+i\eta}G_{\downarrow\downarrow,\downarrow\downarrow} \ , 
\label{eq:gmat}
\end{eqnarray}

\[\left(\begin{array}{cc}
G_{\uparrow\downarrow,\uparrow\downarrow} &
G_{\uparrow\downarrow,\downarrow\uparrow} \\
G_{\downarrow\uparrow,\uparrow\downarrow} &  
G_{\downarrow\uparrow,\downarrow\uparrow}
        \end{array}
  \right) 
=\left(\begin{array}{cc}
V_{\uparrow\downarrow,\uparrow\downarrow} &
V_{\uparrow\downarrow,\downarrow\uparrow} \\
V_{\downarrow\uparrow,\uparrow\downarrow} &  
V_{\downarrow\uparrow,\downarrow\uparrow}
        \end{array}
  \right)
+\left(\begin{array}{cc}
V_{\uparrow\downarrow,\uparrow\downarrow}
&
V_{\uparrow\downarrow,\downarrow\uparrow} 
\\
V_{\downarrow\uparrow,\uparrow\downarrow} 
& 
V_{\downarrow\uparrow,\downarrow\uparrow}
        \end{array}
  \right) \! \!
\left(\begin{array}{cc}
\frac{Q_{\uparrow\downarrow,\uparrow\downarrow}}{\omega-\epsilon_{\uparrow}-\epsilon_{\downarrow}+i\eta}
&
0 
\\
0 &
\frac{Q_{\downarrow\uparrow,\downarrow\uparrow}}{\omega-\epsilon_{\downarrow}-\epsilon_{\uparrow}+i\eta} 
        \end{array}
  \right) \! \!
\left(\begin{array}{cc}
G_{\uparrow\downarrow,\uparrow\downarrow} &
G_{\uparrow\downarrow,\downarrow\uparrow} \\
G_{\downarrow\uparrow,\uparrow\downarrow} &  
G_{\downarrow\uparrow,\downarrow\uparrow}
        \end{array}
  \right) \] \,
 
%\begin{equation}
%G^{\alpha\rightarrow\beta}=V^{\alpha\rightarrow\beta}+\sum_{\delta}V^{\alpha\rightarrow\delta}\frac{Q^{\delta\rightarrow\delta}}{\omega-E_{\delta}+i\eta}
%G^{\delta\rightarrow\beta} \ .
%\label{eq:gmat}
%\end{equation}

\normalsize

In the above expressions the first (last) two subindices indicate the spin projections of the
two neutrons in the initial (final) state, $V$ is the bare nucleon-nucleon interaction, $Q$
is the Pauli operator which allows only intermediate states compatible with the Pauli
principle, and $\omega$ is the starting energy defined as  the sum of single--particle 
energies,
$\epsilon_{\uparrow(\downarrow)}$, of the interacting neutrons. Note that 
$G_{\uparrow\downarrow,\uparrow\downarrow}$ and $G_{\downarrow\uparrow,\downarrow\uparrow}$
are obtained from a coupled channel equation due to the mixing
induced by the
interaction.     One can equivalently solve
the Bethe--Goldstone equation in the spin--coupled basis, where 
the interaction is diagonal, although in that case the Pauli operator is
non-diagonal. However, 
the conventional angle-average of the Pauli operator makes it diagonal,
thus reducing the problem to an uncoupled one in each total spin channel.

The single--particle energy of a neutron with momentum $k$ and spin projection 
$\sigma=\uparrow$($\downarrow$) is given by
\begin{equation}
\epsilon_{\sigma}=\frac{\hbar^2k^2}{2m}+U_{\sigma}(k) \ ,
\label{eq:spe}
\end{equation}
where the single--particle potential $U_{\sigma}(k)$ represents the
average field felt by 
the neutron due to its interaction with the other neutrons  of
the system. In the BHF approximation it is given by
\begin{equation}
U_{\sigma}(k)=\mbox{Re}\sum_{\sigma'=\uparrow,\downarrow}\sum_{k' \leq k_F^{\sigma'}}
\langle \vec{k}\vec{k'} |G_{\sigma\sigma',\sigma\sigma'}(\omega=\epsilon_{\sigma}+
\epsilon_{\sigma'})| \vec{k}\vec{k}'\rangle_{\cal A}  \ ,
\label{eq:spp}
\end{equation}
where a sum over the two Fermi seas of spin up and down, characterized by 
$k_F^{\uparrow}=(6\pi^2\rho^{\uparrow})^{1/3}$ and  
$k_F^{\downarrow}=(6\pi^2\rho^{\downarrow})^{1/3}$ respectively, is performed and the 
matrix elements are properly antisymmetrized.

Once a self-consistent solution of Eqs. (\ref{eq:gmat}) and (\ref{eq:spp}) is obtained, 
the total energy per particle is easily calculated 
\begin{equation}
\frac{E}{N}=\sum_{\sigma=\uparrow,\downarrow}\sum_{k \leq k_F^{\sigma}}\left(\frac{\hbar^2k^2}{2m}+\frac{1}{2}U_{\sigma}(k)\right) \ .
\label{eq:epp}
\end{equation}
This quantity is a function of $\rho^{\uparrow}$ and $\rho^{\downarrow}$ or, equivalently, 
of the total density $\rho=\rho^{\uparrow}+\rho^{\downarrow}$ and the spin polarization
$\Delta$, defined as
\begin{equation}
\Delta=\frac{\rho^{\uparrow}-\rho^{\downarrow}}{\rho} \ .
\label{eq:spin_pol}
\end{equation}
Note that the value $\Delta=0$ corresponds to non-polarized or paramagnetic 
($\rho^{\uparrow}=\rho^{\downarrow}$) neutron matter, whereas $\Delta=\pm 1$ means that the
system is totally polarized, i.e., all the spins are aligned in the same
direction.

\subsection{Magnetic susceptibility}
\label{sec:sec2.2}

The magnetic susceptibility of a system characterizes the response of this system to 
a magnetic field and gives a measure of the energy required to produce a net spin
alignment in the direction of the field. It is defined as
\begin{equation}
\chi=\left(\frac{\partial{\cal M}}{\partial{\cal H}}\right)_{{\cal H}=0} \
,
\label{eq:suscept1}
\end{equation}
where ${\cal M}$ is the magnetization of the system per unit volume given by
\begin{equation}
{\cal M}=\mu(\rho^{\uparrow}-\rho^{\downarrow})=\mu\rho\Delta \ ,
\label{eq:magneti}
\end{equation}
with $\mu$ the magnetic moment of a neutron, and ${\cal H}$ is the magnetic field 
which can be obtained from
\begin{equation}   
{\cal H}=\rho\left (\frac{\partial (E/N)}{\partial {\cal M}}\right )_{{\cal
M}=0}
=\frac{1}{\mu}\left (\frac{\partial (E/N)}{\partial \Delta} \right )_{\Delta=0}\ .
\label{eq:field}
\end{equation}  

Using Eqs. (\ref{eq:magneti}) and (\ref{eq:field}), the magnetic
susceptibility can be written as
\begin{equation}
\chi=\frac{\mu^2\rho}{\left(\frac{\partial^2(E/N)}{\partial\Delta^2}\right)_{\Delta=0}} \ ,
\label{eq:suscept2}
\end{equation}
where the second derivative can be taken at $\Delta=0$ if the field ${\cal H}$ is assumed 
to be small.

It is customary to study the magnetic susceptibility in terms of the ratio $\chi/\chi_F$, 
where $\chi_F$ is the magnetic susceptibility of a free Fermi gas, usually known as
Pauli susceptibility. It can be straightforwardly obtained from Eq. (\ref{eq:suscept2}) and
the total energy per particle of the free Fermi gas
\begin{equation}
\chi_F=\frac{\mu^2m}{\hbar^2\pi^2}k_F \ .
\label{eq:suscept3}
\end{equation}
where the Fermi momentum $k_F=(3\pi^2\rho)^{1/3}$ is related to $k_F^{\uparrow}$ and
$k_F^{\downarrow}$ through the relations
\begin{equation}
\begin{array}{c}
k_F^{\uparrow}=k_F(1+\Delta)^{1/3} \nonumber \\
k_F^{\downarrow}=k_F(1-\Delta)^{1/3} \ .
\end{array}
\label{eq:fermi_mon}
\end{equation}

%%%%%%%%%%%%%%%%%%%%%%%%%%%%%%%%%%%%%%%%%%%%%%%%%%%%%%%%%%%%%%%%%%%%%%%%%%%%%%%%%%%%%%%%%%%%%%%%%%%%%%%%%%%%%%%%%%%%

\section{Results}
\label{sec:sec3}

\begin{figure}[hbtp]
 \setlength{\unitlength}{1mm}
       \begin{picture}(100,180)
       \put(15,10){\epsfxsize=12cm \epsfbox{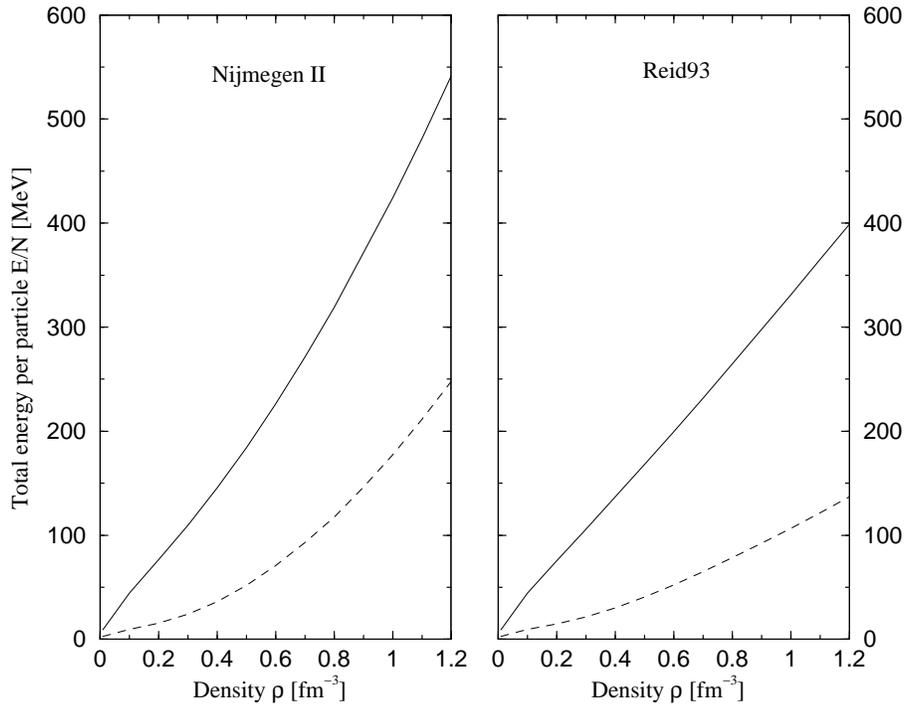}}
       \end{picture}
   \caption{Total energy per particle as a function of the density for totally polarized
(solid lines) and non-polarized (dashed lines) neutron
matter. The left panel shows results for the Nijmegen II nucleon-nucleon interaction, 
whereas results on the right panel correspond to the Reid93 interaction.}
   \label{fig:fig1}
\end{figure}

\begin{figure}[hbtp]
 \setlength{\unitlength}{1mm}
       \begin{picture}(100,180)
       \put(15,10){\epsfxsize=12cm \epsfbox{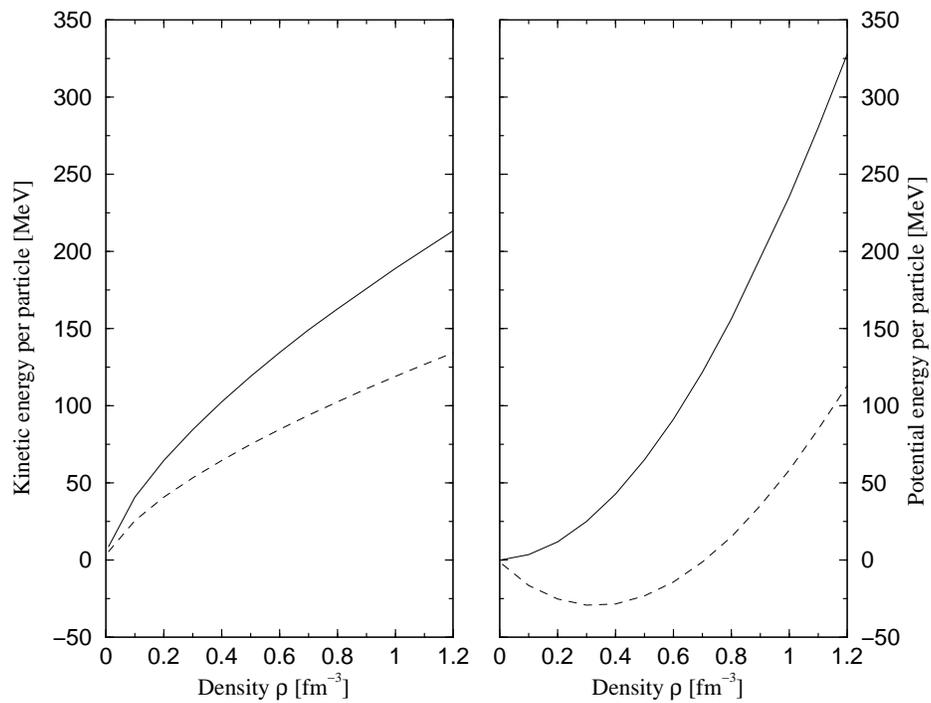}}
       \end{picture}
   \caption{Kinetic (left panel) and potential (right panel) energy contributions to the
total energy per particle as a function of the density for totally
polarized (solid lines) and non-polarized (dashed lines) neutron matter. Results are shown for the Nijmegen II interaction. 
}
   \label{fig:fig2}
\end{figure}

The total energy per particle 
 for totally polarized (solid lines)
and non-polarized (dashed lines) neutron matter is shown in Fig.\ \ref{fig:fig1}
as a function of the density.
 Results for 
the Nijmegen II interaction are plotted on the left panel, whereas  those corresponding 
to the Reid93 interaction are shown on the right panel. As can be seen from the figure, for
both interaction models, the energy of totally polarized neutron matter is always more repulsive than
non-polarized neutron matter in all the density range explored. This additional repulsion 
can be understood, firstly, in terms of the kinetic energy contribution, which is larger in
the totally polarized case than in the non-polarized one; and secondly, in terms of the 
potential energy contribution because, due to symmetry arguments, all
partial waves with
even orbital angular momentum $L$ (some of them attractive, as the important $^{1}S_0$) are
excluded in totally polarized neutron matter.  In order to illustrate this, we have plotted 
 in Fig.\ \ref{fig:fig2} the separate kinetic (left panel) and potential (right panel) energy
contributions
 for the Nijmegen II interaction model (similar results are obtained for the 
Reid93 one, but they are not included in order to make  the discussion more clear). An
interesting conclusion which can be inferred from these results is that a phase 
transition to a ferromagnetic state is not to be expected from our calculation. If such a
transition would exist a crossing of the energies of the  totally polarized and the
non-polarized cases would be observed at some density, indicating that the ground state 
of the system would be ferromagnetic from that density on. As can be seen in
Fig.\ \ref{fig:fig1}, there is no sign of such a crossing and, on the contrary, it becomes less favourably as the density increases.   

\begin{figure}[hbtp]
 \setlength{\unitlength}{1mm}
       \begin{picture}(100,180)
       \put(15,10){\epsfxsize=12cm \epsfbox{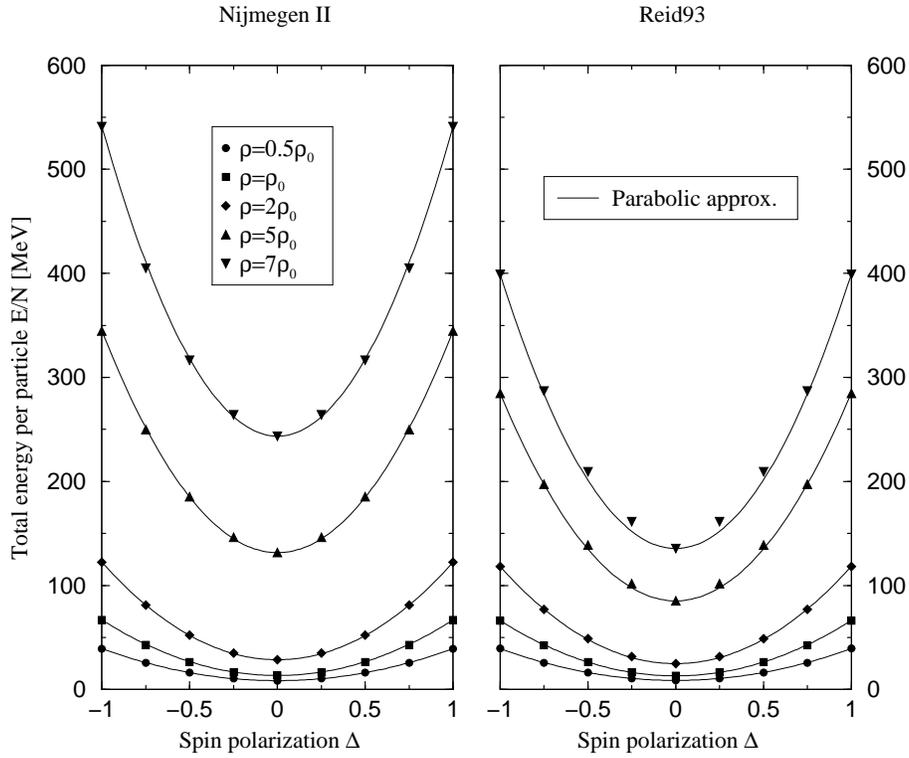}}
       \end{picture}
   \caption{Total energy per particle as a function of the spin polarization $\Delta$ for
different densities. Results for the Nijmegen II (Reid93)
interaction are shown in the left (right) panel. Circles, squares, diamonds and triangles show  our BHF results, whereas solid lines correspond to the
parabolic approximation defined in Eq. (\ref{eq:parabol}).}
   \label{fig:fig3}
\end{figure}

We have shown results for totally polarized and non-polarized neutron matter. Let us consider 
now an intermediate situation in which not all the spins, but a part of them, are aligned
in a given direction, and let us examine the dependence of the total energy per particle in
the spin polarization $\Delta$. This dependence is shown in Fig.\ \ref{fig:fig3}
for five different densities
($\rho_0/2, \rho_0, 2\rho_0, 5\rho_0$ and $7\rho_0$, being $\rho_0=0.17$ fm$^{-3}$ the 
saturation density of nuclear matter). As in Fig.\ \ref{fig:fig1},
results for the Nijmegen (Reid93) are shown on the left (right) panel. Circles, squares,
diamonds and triangles correspond to our BHF results, whereas solid lines correspond to the
parabolic approximation discussed below. As can be seen from this figure, and as it was
expected, $E/N$ is symmetric in $\Delta$. It can also be seen in this
figure that $E/N$
shows a minimum at $\Delta=0$ for all the densities considered, being this again an
indication that the ground state of neutron matter is paramagnetic. Another interesting
thing is to note that this dependence is up to a very good approximation parabolic, being
this parabolic character only slightly lost at very large  densities. Therefore, in
the same spirit as it is done in nuclear matter to determine the symmetry energy, one
can try to
characterize this dependence in the following simple analytic form
\begin{equation}
\frac{E}{N}(\rho,\Delta)=\frac{E}{N}(\rho,0)+a(\rho)\Delta^2 \ ,
\label{eq:parabol}
\end{equation}
where, assuming the quadratic dependence to be valid up to $\mid \Delta\mid = 1$
as our results indicate, the value 
of $a(\rho)$ can be easily obtained for each density as the difference between the total
energy per particle of totally polarized and non-polarized neutron matter
\begin{equation}
a(\rho)=\frac{E}{N}(\rho,\pm 1) - \frac{E}{N}(\rho,0) \ .
\label{eq:valuea}
\end{equation}

The magnetic susceptibility can be evaluated in a very simple way if the parabolic dependence 
of Eq. (\ref{eq:parabol}) is assumed, giving
\begin{equation}
\chi(\rho)=\frac{\mu^2\rho}{2a(\rho)} \ .
\label{eq:sus4}
\end{equation}

\begin{figure}[hbtp]
 \setlength{\unitlength}{1mm}
       \begin{picture}(100,180)
       \put(15,10){\epsfxsize=12cm \epsfbox{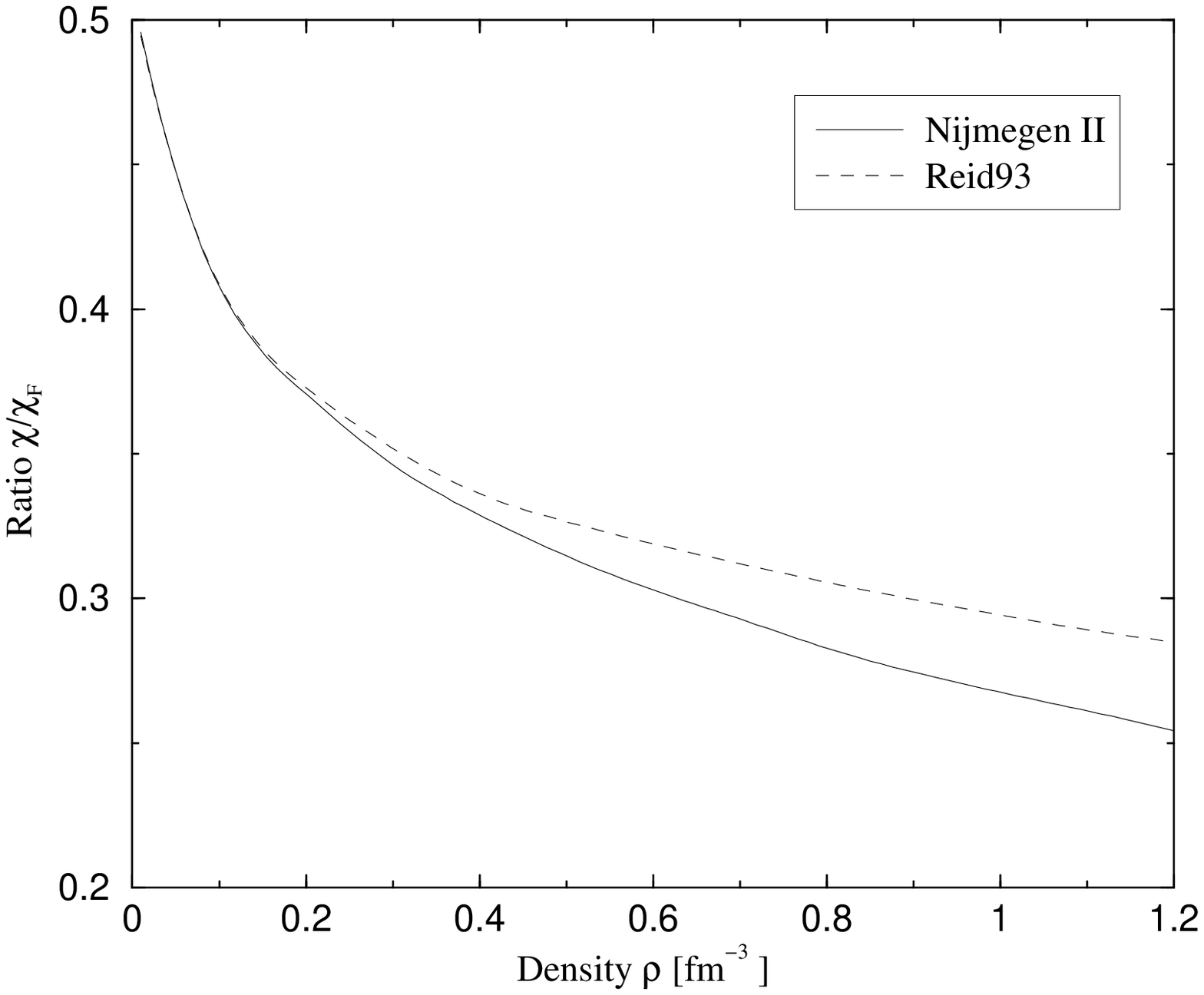}}
       \end{picture}
   \caption{Ratio $\chi/\chi_F$ as a function of the density. The solid line shows the
result for the Nijmegen II interaction, while the dashed line
corresponds to the one obtained with Reid93.}
   \label{fig:fig4}
\end{figure}

In Fig.\ \ref{fig:fig4} the ratio $\chi/\chi_F$ is shown as a function of the density. The
solid line shows the result for the Nijmegen II interaction, whereas the
dashed line corresponds
to the one obtained with Reid93. Starting from 1, the  ratio decreases rapidly  for small 
densities and more slowly as density increases. It can be inferred again from this
figure that a ferromagnetic phase transition, which would be signaled by
an infinite discontinuity
giving rise to a change of sign in $\chi/\chi_F$, is not seen and not
expected at larger
densities either. 

Finally, our results for $\chi/\chi_F$ are compared in Table \ref{tab:tab1} with those 
of the recent calculation performed by Fantoni {\it et al.} \cite{fa01}, shown in columns  
labelled AU6' and AU8'. As can be seen from the table, there is a very good agreement
between these results and ours. For completeness, we show in parentheses  the results
obtained when the standard discontinuous prescription is used in solving the Bethe--Goldstone
equation. In both prescriptions, the results are very similar, which is not surprising
 due to the fact that $\chi$ is obtained from an energy difference 
(see Eqs. (\ref{eq:valuea}) and(\ref{eq:sus4})) which partly cancels
the possible discrepancies. Only for densities larger than $2 \rho_0$, the discontinuous
prescription results differ more than $10 \%$ from the continuous ones.

%%%%%%%%%%%%%%%%%%%%%%%%%%%%%%%%%%%%%%%%%%%%%%%%%%%%%%%%%%%%%%%%%%%%%%%%%%%%%%%%%%%%%%%%%%%%%%%%%%%%%%%%%%%%%%%%%%%%%%%%%%%%%%%%%%%%%%%%%%%%%%%%%%%%%%%%%%%%%

\section{Summary and Conclusions}
\label{sec:sec4}

Employing realistic modern nucleon-nucleon interactions (Nijmegen II and Reid93) we have performed 
a Brueckner--Hartree--Fock calculation of spin polarized neutron matter
properties. We have studied the total energy per particle of neutron matter as a function
of the density and the spin polarization $\Delta$. We have found that in  the range of
densities explored (up to $7\rho_0$) totally polarized matter is always more repulsive than
non-polarized matter, being this an indication that a phase transition of the system to a 
ferromagnetic state is not expected.

We have seen that the total energy per particle is not only symmetric on the spin
polarization $\Delta$, as it was expected, but also parabolic in a very good approximation
 up to $\mid \Delta \mid =1$ even at high densities. This finding supports the calculation
of the magnetic susceptibility of neutron matter by using only the energies of the 
spin symmetric and fully polarized systems. 

Finally, we have calculated  the magnetic susceptibility of the system as a function of the
density, finding a very good agreement with a recent Monte Carlo
calculation \cite{fa01}.

%%%%%%%%%%%%%%%%%%%%%%%%%%%%%%%%%%%%%%%%%%%%%%%%%%%%%%%%%%%%%%%%%%%%%%%%%%%%%%%%%%%%%%%%%%%%%%%%%%%%%%%%%%%%%%%%%%%%%%%%%%%%%%%%%%%%%%%%%%%%%%%%%%%%%%%%%%%%%

\section*{Acknowledgements}
This work is partially supported by the DGICYT contract No. PB98-1247
and by the Generalitat de Catalunya grant No. 2000SGR00024 (Spain). The authors
acknowledge useful discussions with Prof. J. Navarro. One of  
the authors (I.V.) wishes to acknowledge support from a post-doctoral fellowship
of the Istituto Nazionale di Fisica Nucleare (Italy).

%%%%%%%%%%%%%%%%%%%%%%%%%%%%%%%%%%%%%%%%%%%%%%%%%%%%%%%%%%%%%%%%%%%%%%%%%%%%%%%%%%%%%%%%%%%%%%%%%%%%%%%%%%%%%%%%%%%%%%%%%%%%%%%%%%%%%%%%%%%%%%%%%%%%%%%%%%%%%

%%%%%%%%%%%%%%%%%%%%%%%%%%%%%%%%%%%%%%%%%%%%%%%%%%%%%%%%%%%%%%%%%%%%%%%%%%%%%%%%%%%%%%%%%%%%%%%%%%%%%%%%%%%%%%%%%%%%%%%%%%%%%%%%%%%%%%%%%%%%%%%%%%%%%%%%%%%%%

\begin{table}
\caption{Magnetic susceptibility ratio $\chi/\chi_F$. Our BHF results, 
labelled Nijmegen II and Reid93, are compared with the AFDMC results of Fantoni {\it et
al.} [12], labelled AU6' and AU8'. BHF results obtained with the standard
discontinuous prescription are
given in parentheses.} 
\bigskip
\bigskip
\begin{tabular}{c| cc| cc}
\phantom{cac} $\rho/\rho_0$ \phantom{cac} & Nijmegen II & Reid93 \phantom{cac} & AU6' & AU8' \phantom{cac} \cr
\hline
\phantom{cac} $0.75$ \phantom{cac} & $0.39$($0.41$)  & $0.39$($0.41$) \phantom{cac} & $0.40$ & \phantom{cac} \cr
\phantom{cac} $1.25$ \phantom{cac} & $0.38$($0.39$)  & $0.37$($0.39$) \phantom{cac} & $0.37$ & $0.39$ \phantom{cac} \cr
\phantom{cac} $2.0$  \phantom{cac} & $0.34$($0.37$)  & $0.34$($0.38$) \phantom{cac} & $0.33$ & $0.35$ \phantom{cac} \cr
\phantom{cac} $2.5$  \phantom{cac} & $0.32$($0.36$)  & $0.33$($0.37$) \phantom{cac} & $0.30$ & \phantom{cac} \cr
\end{tabular}
\label{tab:tab1}
\end{table}

%\begin{table}
%\caption{Magnetic susceptibility ratio $\chi/\chi_F$. Our BHF results are compared with those of Fantoni and Sarsa.}
%\bigskip
%\bigskip
%\begin{tabular}{c| cc |cc}
%$\rho/\rho_0$ \phantom{cac}& Nijmegen II & Reid93  \phantom{caca}& AU6' & AU8' \cr
%\hline
%$0.75$ \phantom{cac} & $0.39$  & $0.39$  \phantom{cac}& $0.39$ & $$ \cr
%$1.25$ \phantom{cac} & $0.38$  & $0.37$  \phantom{cac}& $0.37$ & $0.39$ \cr
%$2.0$ \phantom{cac} & $0.34$  & $0.34$  \phantom{cac}& $0.32$ & $0.35$ \cr
%$2.5$ \phantom{cac} & $0.32$  & $0.33$  \phantom{cac}& $0.31$ & $$ \cr
%\end{tabular}
%\label{tab:tab1}
%\end{table}

%%%%%%%%%%%%%%%%%%%%%%%%%%%%%%%%%%%%%%%%%%%%%%%%%%%%%%%%%%%%%%%%%%%%%%%%%%%%%%%%%%%%%%%%%%%%%%%%%%%%%%%%%%%%%%%%%%%%%%%%%%%%%%%%%%%%%%%%%%%%%%%%%%%%%%%%%%%%%

\begin{thebibliography}{200}

\bibitem{go68} T. Gold, Nature {\bf 218}, 731 (1968).

\bibitem{br69} D. H. Brownell Jr. and J. Callaway, Nuovo Cimento {\bf 60} B, 169 (1969).

\bibitem{ri69} M. J. Rice, Phys. Lett. {\bf 29A} , 637 (1969).

\bibitem{si63} S. D. Silverstein, Phys. Rev. Lett. {\bf 23}, 139 (1963).

\bibitem{os70} E. \O stgaard, Nucl. Phys. {\bf A154}, 202 (1970).

\bibitem{cl69} J. W. Clark, Phys. Rev. Lett. {\bf 23}, 1463 (1969).

\bibitem{pe70} J. M. Pearson and G. Saunier, Phys. Rev. Lett. {\bf 24}, 325 (1970).

\bibitem{pa72} V. R. Pandharipande, V. K. Garde and J. K. Srivastava,
Phys. Lett. {\bf 38B}, 485 (1972).

\bibitem{ba73} S. O. B\"ackmann and C. G. K\"allman, Phys. Lett. {\bf
43B}, 263 (1973).

\bibitem{ja82} A. D. Jackson, E. Krotscheck, D. E. Meltzer and
R. A. Smith, Nucl. Phys. {\bf A386}, 125 (1982).

\bibitem{vi84} A. Vidaurre, J. Navarro and J. Bernab\'eu, Astron. Astrophys. {\bf 135}, 361 (1984).

\bibitem{fa01} S. Fantoni, A. Sarsa and E. Schmidt, Phys. Rev. Lett. {\bf 87}, 181101 (2001).

\bibitem{sm97} A. Smerzi, D. G. Ravenhall and V. R. Pandharipande, Phys. Rev. C {\bf 56}, 
2549 (1997).

\bibitem{pi98} S. C. Pieper, in {\it Microscopic Quantum Many-Body Theories and their
Applications}. Lecture Notes in Physics {\bf 510}, 337 (Springer Verlag, Berlin, 1998).

\bibitem{so98} H. Q. Song, M. Baldo, G. Giansiracusa and U. Lombardo, Phys. Rev. Lett. 
{\bf 81}, 1585 (1998).

\bibitem{st94} V. G. J. Stoks, R. A. M. Klomp, C. P. F. Terheggen and L. J. de Swart, Phys. Rev. C {\bf 49}, 2950 (1994).

\end{thebibliography}
\end{document}